\documentstyle[12pt]{article}
\newcommand{\beq}{\begin{equation}}
\newcommand{\eeq}{\end{equation}}

\def\op{operator}

\def\cn{condition}
\def\LG{Landau-Ginzburg}
\def\sg2{\int_{\Sigma_2}\sqrt{g^{(2)}}}
\def\sf{\int_{\Sigma_1}\sqrt{g^{(1)}}}

\begin{document}
\begin{titlepage}
\begin{flushright}
NBI-HE-94-35 \\
July 1994\\
\end{flushright}
\vspace{2.5cm}
\begin{center}
{\Large {\bf Realization of $W_{1+\infty}$ and Virasoro Algebras\\
in Supersymmetric Theories on Four Manifolds}}\\
\vspace{1cm}
\end{center}
\begin{center}
{\large Andrei\ Johansen}\footnote{E-mail: ajohansen@nbivax.nbi.dk \ / \
johansen@lnpi.spb.su \ }
\\ \mbox{} \\
{\it The Niels Bohr Institute,}\\
{\it Blegdamsvej 17, 2100 Copenhagen, Denmark}\\ \vskip .2 cm
and \\ \vskip .2 cm
{\it The St.Petersburg Nuclear Physics Institute,}\\
{\it Gatchina, 188350 St.Petersburg, Russian Federation}\\
\end{center}
\begin{abstract}

We demonstrate
that a supersymmetric theory twisted on a K\"ahler four manifold
$M=\Sigma_1 \times \Sigma_2 ,$ where $\Sigma_{1,2}$ are 2D Riemann
surfaces,
possesses a "left-moving" conformal stress tensor on $\Sigma_1$
($\Sigma_2$) in the BRST cohomology.
The central charge of the Virasoro algebra has a purely geometric
origin and is proportional to the Euler characteristic of the $\Sigma_2$
($\Sigma_1$) surface.
This structure is shown to be invariant under renormalization group.
We also give a representation of the algebra $W_{1+\infty}$
in terms of a free chiral supermultiplet.

\end{abstract}
\end{titlepage}
\newpage

\section{Introduction}
\setcounter{equation}{0}

Recently \cite{ajoh} it has been shown that four dimensional
$N=1$ SUSY gauge
theories with an appropriate representation of matter can be twisted
on a K\"ahler manifold.
A twisted supersymmetric gauge
theory possesses a BRST \op \ which is nilpotent
and does not depend on the external metric.
If there is no superpotential for the matter supermultiplet then
the action of the theory turns out to be BRST exact.
Otherwise it is BRST closed.
The topological Yang-Mills theory \cite{witten} is actually a
particular example of such a twisted supersymmetric gauge
model with the fields of matter in
the adjoint representation of the gauge group.
However in the topological Yang-Mills theory
\cite{witten} and in its twisted $N=1$ version
one considers the cohomologies of the different BRST \op s
\footnote{The formulation of Witten's theory in terms of N=1 twisted
SUSY turns out to be very useful for calculations of the Donaldson
invariants \cite{witten4}.}.
An extension of this construction to the supersymmetric theories
without gauge interactions is straightforward (we illustrate it in the
present paper).
We will henceforth generically name the supersymmetric theories twisted
on a K\"ahler manifold heterotic topological theories.

Such a twisting of the $N=1$ supersymmetric models
is a four dimensional analogue of a half-twisting
of the 2D theories \cite{WiLG}.
Witten \cite{WiLG} has shown that under the operation of a half-twist the
\LG \ model turns out to be left-moving and conformally invariant.
This enables us to extract an essential information about the $N=2$
superconformal algebra realized at the infrared fixed point of the \LG \
model.
In particular Witten computed the corresponding elliptic genus.
This superconformal algebra is realized on the classes of cohomology
of one of the supergenerators of the $N=2$ SUSY algebra.
Witten also extended his analysis to the case of $(0,2)$ SUSY models
coupled to abelian gauge multiplets \cite{SiWi}.
In Ref.\cite{mohri} this construction has been extended to more
complicated 2D $N=2$ models where $W_N$ algebras are realized in
the cohomology of one of the supergenerators.

The space of physical \op s (defined as the classes of cohomology of an
appropriate BRST \op ) in the heterotic theory is in general
more complicated than that of the topological theory.
This space contains a ground ring of the local operators which are BRST
invariant off-shell.
This is a purely topological part of the model and this ring coincides
with the ring of local physical \op s in Witten's topological Yang-Mills
theory provided that
the matter is in the adjoint representation of the gauge group.
Moreover there exist also \op s in the cohomology
of the BRST \op \ which can depend on the external metric and commute with
the BRST \op \ only up to the equations of motion.
This structure of the space of physical \op s is similar to that of
2D half-twisted models \cite{WiLG,SiWi}.
In general the physical correlators after the insertions of
such \op s depend on the external metric.
This is due to the presence of such a metric in the inserted \op s.
They can also depend on the holomorphic coordinates of the K\"ahler
manifold but not on the anti-holomorphic ones \cite{ajoh}.
In the theory without a superpotential
the action is BRST exact and, hence, the physical correlators do not
depend on the gauge coupling constant.
Therefore the path integral for the
physical correlators allows for a localization near the solution of the
classical equations of motion.

In the present paper \footnote{It is a short version of ref.\cite{joh}.}
we show that the heterotic topological
theory on a four manifold $M=\Sigma_1 \times \Sigma_2$, where
$\Sigma_{1,2}$ are 2D Riemann surfaces, possesses two chiral Virasoro
algebras in the cohomology of a BRST \op .
The central charges of these algebras are invariant under the
renormalization group and can be calculated in the weak coupling limit.
These central charges turn
out to be proportional to the Euler characteristic of the corresponding
Riemann surface ($\Sigma_1$ or $\Sigma_2$).
One can hope that such a construction can provide us with
an important information on the renormalization group in
the four dimensional $N=1$
supersymmetric QCD, or it can be useful in the description of string vacua.

We also demonstrate that the $W_{1+\infty}$ \cite{wintro,pope}
algebra can be realized in terms of a free chiral supermultiplet.

The paper is organized as follows.
In section 2 we formulate the heterotic topological theory.
In section 3 the $W_{1+\infty}$ algebra is realized in terms of a free
chiral supermultiplet.
In section 4 we consider a realization of a Virasoro algebra in the
interacting heterotic topological theories.
We close with a brief summary of our results.

\section{Supersymmetric theory on a four dimensional
K\"ahler manifold}
\setcounter{equation}{0}

The twisted Lagrangian for the Yang-Mills supermultiplet coupled
to a chiral supermultiplet in a representation $R$ of the gauge group
reads as follows \cite{ajoh}
\beq
L =\frac{1}{e^2} \sqrt{g} {\rm Tr} [ F^{\bar{m}\bar{n}} F_{\bar{m}\bar{n}}
+i \bar{\lambda}^{mn} D_m \chi_n -
\frac{1}{2} D'^2 +i g^{\bar{m} m} D' F_{\bar{m} m}
+ i\bar{\lambda} D^m \chi_m] +
\eeq
$$+\sqrt{g} ( \bar{\phi} D^{\bar{m}} D_{\bar{m}} \phi +
\bar{\psi}^{\bar{m}\bar{n}} D_{\bar{m}} \psi_{\bar{n}} + \bar{\psi}
D^{\bar{m}} \psi_{\bar{m}} + N_{\bar{m}\bar{n}}
\bar{N}^{\bar{m}\bar{n}} -$$
$$- i\bar{\phi} \chi_n \psi^n -\frac{1}{2} \bar{\psi} \bar{\lambda}
\phi + \frac{1}{4} \bar{\psi}_{mn} \bar{\lambda}^{mn} \phi + \frac{1}{2}
\bar{\phi} D' \phi )=$$
$$= \frac{1}{e^2} \sqrt{g}
\{ Q,{\rm Tr} [-i\bar{\lambda}  g^{\bar{m} m} F_{\bar{m} m} -
\frac{1}{2} D' \bar{\lambda} -
\frac{i}{2} \bar{\lambda}_{\bar{m}\bar{n}}
F^{\bar{m}\bar{n}}]+$$
$$ +\sqrt{g}
\{ Q,-\frac{1}{2} \bar{\psi}^{\bar{m}\bar{n}}N_{\bar{m}\bar{n}} +
\bar{\phi} D^{\bar{m}} \psi_{\bar{m}} - \frac{1}{2} \bar{\phi}
\bar{\lambda} \phi \} ,$$
where $g$ is a determinant of the external K\"ahler metric
$g_{m\bar{m}}$,
$m$, $n...$ and $\bar{m}$, $\bar{n} ...$ are holomorphic and
anti-holomorphic world indices on a K\"ahler manifold $M,$ respectively;
$D_m$ and $D_{\bar{m}}$ stand for the covariant derivatives in
gravitational and gauge fields.
The gauge supermultiplet consists of
the strength tensor $F_{\mu\nu}$ of the gauge field $A_{\mu} ,$
the fermionic (gaugino \cite{ajoh}) fields $\chi_n ,$ $\bar{\lambda}$ and
$\bar{\lambda}_{\bar{m}\bar{n}}$ which are the $(1,0),$ $(0,0)$ and $(0,2)$
forms respectively, and the $D'$ field which is an auxiliary scalar field.
The chiral supermultiplet includes
the scalar fields $\phi$ and $\bar{\phi} ,$ the fermionic fields
$\psi_{\bar{n}} ,$ $\bar{\psi}$ and $\bar{\psi}_{mn}$ which are $(0,1),$
$(0,0)$ and $(2,0)$ forms respectively, and the $N_{\bar{m}\bar{n}}$ and
$\bar{N}_{mn}$ auxiliary fields.
$Q$ is the scalar nilpotent generator of BRST transformations
which for the gauge multiplet reads
\beq
\delta A_n = \chi_n,\;\;\delta A_{\bar{n}} =0,\;\;
\delta \chi_n =0,
\eeq
$$\delta \bar{\lambda} = - D',\;\;
\delta \bar{\lambda}_{\bar{m} \bar{n}} =
2i F_{\bar{m} \bar{n}},\;\;
\delta D' = 0,$$
while for the chiral multiplet we have
\beq
\delta \phi =0,\;\;\; \delta \psi_{\bar{m}} = D_{\bar{m}}
\phi ,\;\;\; \delta N_{\bar{m}\bar{n}}= D_{\bar{m}}
\psi_{\bar{n}} - D_{\bar{n}}\psi_{\bar{m}} +\frac{1}{2}
\bar{\lambda}_{\bar{m}\bar{n}} \phi ,
\eeq
$$\delta\bar{\phi} =\bar{\psi} ,\;\;\;
\delta \bar{\psi} =0 ,\;\;\;
\delta \bar{\psi}^{\bar{m}\bar{n}} = -2 \bar{N}^{\bar{m}\bar{n}} ,\;\;\;
\delta \bar{N}^{\bar{m}\bar{n}} =0 .$$
It is easy to see that the generator of these transformations does not
depend on the external metric.

One can see that all the fields are combined into different
supermultiplets.
Indeed in the gauge sector we have the following supermultiplets:
($A_n ,\; \chi_n$), ($\bar{\lambda},\; D'$) and
($\bar{\lambda}_{\bar{m} \bar{n}},\; F_{\bar{m} \bar{n}}$).
In the matter sector the multiplets are given by $(\phi ,\psi_{\bar{m}} ,
N_{\bar{m}\bar{n}}) ,$
$(\bar{\phi} ,\bar{\psi}) ,$ and $(\bar{\psi}_{mn} , \bar{N}_{mn} ).$

Fixing the ghost number of the BRST charge to be 1 we have the following
dimensions ($d$) and ghost numbers ($G$) for the different fields:
\beq
(d,G)(A_n) = (d,G)(A_{\bar{n}}) = (1,0), \;\;
(d,G)(\chi_n) = (1,1)\;\;
\eeq
$$(d,G)(\bar{\lambda}) = (d,G)(\bar{\lambda}_{\bar{m} \bar{n}})=
(2,-1) ,\;\;
(d,G)(D') = (2,0).$$
and
\beq
(d,G) (\phi) = (0,2),\;\;\; (d,G)(\bar{\phi})=(2,-2),\;\;\;
(d,G)(\psi_{\bar{m}}) =(1,1),
\eeq
$$(d,G)(\bar{\psi}) = (d,G)(\bar{\psi}_{mn}) =(2,-1),\;\;\;
(d,G)(N_{\bar{m}\bar{n}})=(2,0),\;\;\; (d,G)(\bar{N}_{mn})=(2,0) .$$

Let us now consider a model of a chiral supermultiplet
without gauge interactions.
The Lagrangian of a free chiral supermultiplet can be easily read off
from eq.(2.1).
If there is a nontrivial holomorphic $(2,0)$ form $E_{mn}$ on $M$
(i.e. $H^{2,0} (M) \neq 0$) then it is
possible to introduce a superpotential $W(x)$ which induces masses
and self-interactions for the quantum fields as follows
\beq
L_{int} = E^{\bar{m}\bar{n}} (\psi_{\bar{m}} \psi_{\bar{n}} W'' (\phi) +
N_{\bar{m}\bar{n}} W' (\phi)) -
S^{mn} (\bar{\psi}_{mn}\bar{\psi} W'' (\bar{\phi}) + 2 \bar{N}_{mn} W'
(\bar{\phi}))=
\eeq
$$= E^{\bar{m}\bar{n}} (\psi_{\bar{m}} \psi_{\bar{n}} W'' (\phi) +
N_{\bar{m}\bar{n}} W' (\phi)) + \{ Q, S^{mn} \bar{\psi}_{mn} W'
(\bar{\phi}) \} ,$$
where $S_{\bar{m}\bar{n}}$ is an arbitrary non-singular (0,2) form on $M
.$
It is easy to see that
\beq
\{ Q, \int_M \sqrt{g} L_{int} \} =0,
\eeq
and hence the total action of the theory with a superpotential is
$Q$-closed (but not $Q$-exact due to the interaction terms).

The \cn \ of renormalizability for a four dimensional theory implies that
$W(x)$ should be a polynomial of a degree not higher than 3.
In the case of the superpotential of a degree 2 a Majorana mass
is induced for the
chiral supermultiplet, while for the cubic superpotential the $|\phi|^4$ +
Yukawa interactions are induced.
As it is well known \cite{book} the superpotential in SUSY theories is
non-renormalizable \footnote{Notice however that some subtleties can appear due
to infrared effects \cite{IR}.}.
The only renormalization of the action comes from $D$-terms \cite{book}.
This fact was very important in the analysis of 2D N=2 SUSY theories
\cite{Ma,Va,LVW,howe,cec1,cec2,class,WiLG}, where the $N=2$ superconformal
theories were associated to quasihomogeneous superpotentials.
In an analogy with the 2D case in
this paper we shall consider the case $W(x)=
\lambda x^{N+1}/(N+1) ,$ where $\lambda$ is a coupling constant,
and formally we shall consider all positive integer values of $N.$

We can now define the physical \op s as classes of cohomology of the
BRST \op \ $Q.$
The local observable for the sector of the gauge multiplet
becomes a (2,0) form (of dimension 2) given by
\beq
O^{(0)}_{mn} = Tr \chi_m \chi_n .
\eeq
It is to be remarked that the situation here is different from that in
ordinary topological theories where the local observables are zero-forms;
non-zero forms should usually be integrated over closed cycles
to get non-local observables (in the case of the highest forms one gets
moduli of the topological theory).
The difference here is due to the splitting of four coordinates
into holomorphic and anti-holomorphic ones, so that the (2,0) form
is effectively a scalar with respect to anti-holomorphic derivatives.
We have
\beq
\partial_{\bar{k}} Tr \chi_m \chi_n
= \{ Q,...\} .
\eeq
It follows from this equation that the physical correlators with insertions
of this operator are holomorphic with respect to its coordinate.
The local \op s in the matter sector are given by the
gauge invariant functions of the (dimensionless) scalar field $\phi$
which correspond to flat directions of the classical
moduli space of vacua \cite{seiberg}.
It is also possible to construct non-local operators.

It is worth noticing that we could use the vector field $-V_{\mu}$ for a
twisting of the theory.
Such a modification of the model corresponds to a change $\epsilon ,\eta
\to \eta ,\epsilon .$
The local \op s in this mirror model
are antiholomorphic up to BRST exact \op s (for example, $\partial_n \phi =
\{Q , ...\}$) while their correlators are anti-meromorphic.

In general the twisted supersymmetric
theory has an anomaly in the decoupling of the external
metric from the path integral.
This anomaly originates from the axial anomaly in the twisting fermionic
current \cite{ajoh}.
There are two contributions: the mixed anomaly which depends on both
the gauge field and the gravitational one, and a purely gravitational anomaly.
It has been
shown in ref.\cite{ajoh} that the mixed anomaly is cancelled if
\beq
C_2 (G) -T(R) = C_2 (G) -\sum_i T(R_i) =0.
\eeq
Here $R=\sum_i R_i ,$ $R_i$ are irreducible representations of the gauge
group; $C_2 (G)$ stands for the Casimir \op \
and $T(R_i)$ is Dynkin index of irreducible representation $R_i$ of the
gauge group $G;$
${\rm Tr}_{Ad} t^a t^b = C_2 (G) \delta^{ab}$ and ${\rm Tr}_{R_i}
t^a t^b = T(R_i) \delta^{ab} ,$ where ${\rm Tr}_{Ad}$ and ${\rm
Tr}_{R_i}$ are the traces in the adjoint and $R_i$ representations of the
gauge group respectively; $t^a$ and $t^b$ stand for generators of the
gauge group.
The condition of cancellation of the gravitational anomaly reads
\cite{ajoh}
\beq
\sum_i {\rm dim} \; R_i -{\rm dim} G = 0,
\eeq
where ${\rm dim} G$ and ${\rm dim} R_i$ stand for the dimensions
of the adjoint and $R_i$ representations of the gauge group
respectively.
The \cn \ (2.10) has been analysed (for a different problem) in
ref.\cite{koh}.
{}From that analysis one can easily extract that if the \cn \ (2.10) is
fulfilled then
\beq
\sum_i {\rm dim} \; R_i - {\rm dim} G  \geq 0 .
\eeq
The equality in the above equation is satisfied only for the matter in the
adjoint representation of the gauge group, i.e. for the twisted $N=2$
Yang-Mills theory \cite{witten}.
Notice also that the mixed anomaly is absent in the case of a twisted
model of chiral supermultiplets without gauge interactions.

We shall assume that the mixed anomaly is cancelled.
The purely gravitational anomaly itself does not spoil the BRST invariance of
the theory and the notion of the BRST cohomology \cite{joh}.

The gravitational anomaly for these models has a natural interpretation
as a conformal anomaly of the Virasoro algebra which appears in the
$Q$-cohomology (see below).
Indeed the form of the corresponding term in the effective action can be
extracted from the 4D conformal anomaly in the trace of the
energy-momentum tensor (see, e.g. \cite{duff}) which in our case reads
\beq
<\theta_{\mu}^{\mu}> = ({\rm dim} R-{\rm dim} G)
\frac{1}{3\cdot 128\pi^2} \; ^*R_{\mu\nu\lambda\rho}^*
R^{\mu\nu\lambda\rho} ,
\eeq
where $R_{\mu\nu\lambda\rho}$ is the Riemann tensor.
The variation of the effective action $S_{eff}$ with respect to the metric
$g_{\mu\nu}$ reads
\beq
\delta S_{eff} = ({\rm dim} R -{\rm dim} G)
\frac{1}{6\cdot 128\pi^2}\int_M d^4 x \sqrt{g}
 ^*R_{\mu\nu\lambda\rho}^* R^{\mu\nu\lambda\rho} \delta \log g .
\eeq
For the case $M= \Sigma_1 \times \Sigma_2$ with a block diagonal metric
\beq
g_{\mu\nu} = \left( \begin{array}{cc} g^{(1)}_{ij} & 0 \\ 0&
g^{(2)}_{kl} \end{array}
\right) ,
\eeq
we get for the effective action
\beq
S_{eff}= ({\rm dim} R -{\rm dim} G)\times
\left( \frac{\chi_1}{48\pi^2} \sg2
R^{(2)}\frac{1}{\Delta^{(2)}} R^{(2)} +
\frac{\chi_2}{48\pi^2} \sf
R^{(1)}\frac{1}{\Delta^{(1)}} R^{(1)} \right) ,
\eeq
where $\chi_{1,2}$ are the Euler characteristics of $\Sigma_{1,2} ,$
$R^{(1)},$ $R^{(2)}$ are the Riemann curvatures of $\Sigma_{1,2} ,$ and
$\Delta^{(1)} ,$ $\Delta^{(2)}$ are the corresponding Laplace \op s.
We observe that this is a sum of two induced Liouville actions for the
conformal theories on $\Sigma_2$ and $\Sigma_1$ with the central charges
$\chi_1$ and $\chi_2 .$
If the gauge interactions are absent the factor $({\rm dim} R -
{\rm dim} G)$ changes into $N_f$ (the number of chiral supermultiplets).

The effective action (2.15) is non-renormalizable at the multi-loop
level.
Indeed from the point of view of the supersymmetry this anomaly
originates from the anomaly in the axial fermionic
current which is used for the twisting of the theory \cite{ajoh}.
Since this current has an anomaly that depends only on the classical
external gravitational
fields it does not have any anomalous dimension and hence
the anomaly is non-renormalizable.
A different argument can be given
which is based on the background superfield formalism for
supergravity coupled to matter supermultiplets \cite{marcgrav}.
Indeed as it has been demonstrated in ref.\cite{marcgrav}, a superfield diagram
for multiloop corrections to an effective action can be represented as
an integral over four Grassmann variables with an integrand which is a
local expression with respect to the Grassmann coordinates.
If we formulate our theory in terms of usual fields (with non-changed
spins) the quantum supermultiplets are coupled to a reduced external
gravitational superfield that is invariant under one of four
supercharges (this supergravity multiplet includes only a
veirbein field and a vector one).
That means that the integrands in superfield diagrams in such a
background do not depend on the Grassmann coordinate corresponding to
that supercharge.
Therefore such multiloop corrections to the effective action in this
special supergravity background vanish due to an integration over
Grassmann coordinates.
It is straightforward to formulate this conclusion in terms of twisted
fields.
Thus we see that there are no multiloop corrections to the gravitational
anomaly (2.15)and (2.17).
This argument is an extension of the usual theorems on
non-renormalizability of a superpotential and of the effective action
in the instanton background \cite{grisaru,vainshtein} to the case of
a special supergravity background.

\section{Realization of the $W_{1+\infty}$ algebra
in terms of a free chiral supermultiplet}
\setcounter{equation}{0}

For definiteness we hereby describe the conformal algebra on
the $\Sigma_1$ Riemann surface.
Let for simplicity $\Sigma_1=T^2$ be a torus with a flat two-dimensional
metric.
The metric of $M=\Sigma_1 \times \Sigma_2$ can be chosen block diagonal
(see eq.(2.16)).

Let us define the following \op s
\beq
W_{n+1} = 2\pi \int_{\Sigma_2} \sqrt{g^{(2)}} ( -\partial_1 \bar{\phi}
\partial^n_1 \phi + g^{2\bar{2}}\bar{\psi}_{12} \partial^n_1
\psi_{\bar{2}} ).
\eeq
Using the definition of $Q$ in eqs.(2.3) and the equations of motion for
a free chiral supermultiplet
it is easy to see that these \op s are $Q$-closed on mass-shell
\beq
\{ Q, W_{n+1} \} = 0 .
\eeq
Moreover these \op s are holomorphic on $\Sigma_1$ in cohomology of $Q$
\beq
\partial_{\bar{1}} W_{n+1} =
2\pi \{ Q, \sg2 ( -\partial_1 \bar{\phi} \partial_1^n \psi_{\bar{1}}
- \partial_2 \bar{\phi} \partial^n_1 \psi_{\bar{2}} ) \} .
\eeq
Thus we see that $\partial_{\bar{1}} W_{n+1}$ is $Q$-exact and hence is
trivial in cohomology.

It is worth emphasizing that the \op s $T=W_2$ and $J=W_1$ are the 11- and 1-
components of 2D tensors on $\Sigma_1 .$
These are actually the components of the energy momentum tensor and
the $U(1)$ current integrated over $\Sigma_2$ respectively.
This fact is responsible for holomorphicity of $T$ and $J.$
Indeed the four dimensional current $J_{\mu}$
that corresponds to an unbroken $U(1)$ $R$-symmetry of the action obeys
\beq
\partial_n J^n +\partial^n J_n =0,
\eeq
while for the energy-momentum tensor $\theta_{\mu\nu}$ we have
\beq
\partial^n \theta_{n1} + \partial^{\bar{n}} \theta_{\bar{n} 1} =0 .
\eeq
When integrated over $\Sigma_2$ the terms which are the total
derivatives in $z^2$ and $\bar{z}^{\bar{2}}$ in these equations vanish.
In turn it is easy to check that the components $J_{\bar{1}}$ and
$\theta_{1\bar{1}}$ are $Q$-exact because of the $Q$-exactness of the
action (for example, $\theta_{1\bar{1}}$ is a variation of the action
with respect to the component $g_{1\bar{1}}$ of the metric).

We now proceed to show that the \op s $W_{n+1}$ generate the
$W_{1+\infty}$ algebra in the cohomology of $Q.$
To this end let us consider the operator product expansion (OPE) of the \op s
$W_{n+1} .$
We have
\beq
W_{n+1} (z,\bar{z}) W_{p+1} (w,\bar{w}) =
\frac{n!p!(-1)^{n+1} \chi}{2(z-w)^{n+p+2}} +
\eeq
$$+2\pi \sum^{\infty}_{k=0} \frac{(z-w)^k}{k!} \sg2 d^2 v \left[
\left( \partial_w \bar{\phi} \partial_w^{n+k} \phi - \bar{\psi}_{12}
\partial_w^{n+k} \psi_{\bar{2}} \right) \frac{p! (-1)}{(z-w)^{p+1}}
+\right.$$
$$\left. \left( \partial_w^{k+1} \bar{\phi} \partial_w^p \phi -
\partial_w^k \bar{\psi}_{12}
\partial_w^p \psi_{\bar{2}} \right) \frac{n! (-1)^n}{(z-w)^{n+1}}
\right] +$$
$$+ \; Q{\rm -exact \;\; terms}=$$
$$= \frac{n!p!(-1)^{n+1} \chi}{2(z-w)^{n+p+2}}+
\frac{p+n}{(z-w)^2} W_{p+n} (w,\bar{w}) + \frac{n}{z-w} \partial_w
W_{n+p} + f(W_{n+p-1} ,..., W_1) +$$
$$+{\rm regular \;\; terms} + \; Q{\rm -exact \;\; terms} ,$$
where $\chi = (-1/2\pi) \sg2  \partial_{\bar{2}}
\partial_{\bar{2}} \rho =2(1-g)$
is the Euler characteristic of the Riemann surface $\Sigma_2 ,$
$z$ and $w$ stand for complex coordinates on the surface $\Sigma_1 .$
Here we used that $\partial_{\bar{n}} \phi = \{ Q, \psi_{\bar{n}} \} ,$
$\bar{\psi} =\{ Q,\bar{\phi} \}$ and the equations of motion.
Thus the OPE for $W_{n+1} W_{p+1}$ is holomorphic on $\Sigma_1$ in
$Q$-cohomology.

It is easy to see that the \op \ $T=W_2$ generates the holomorphic Virasoro
algebra with the central charge $\chi$ which is the Euler characteristic of
$\Sigma_2 .$
Moreover one can see that the \op s $W_{n+1}$ generate
the holomorphic $W_{1+\infty}$ algebra.
Indeed by introducing the Fourier modes
\beq
W^n_s = \oint \frac{dz}{2\pi i} z^{n+s} W_{n+1}(z,\bar{z})
\eeq
we get (in the $Q$-cohomology)
\beq
[W^n_s ,W^p_{s'}] = \frac{\chi (-1)^{n+1} n! p!}{2} \cdot
\frac{(s+n)...(s-p)}{(n+p+1)!} \delta_{s+s' ,0} +
(sp -ns') W^{n+p-1}_{s+s'} + R .
\eeq
Here $R$ stands for terms which depend only on $W^k_i$ with $k<n+p-1 .$
The standard $W_{1+\infty}$ commutation relations correspond to $R=0$
\cite{wintro,pope,bakas} while in our algebra $R\neq 0.$
However adding to the \op s $W_{n+1}$ appropriate linear combinations
$\sum_{k=0}^{n-1} a_k \partial_z^{n-k} W_{k+1} ,$ where $a_k$ are
constant coefficients, one can recover the standard
commutation relations with $R=0.$

\section{Interacting fields}

Let us now consider a manifold $M=\Sigma_1 \times \Sigma_2$ where
the genera of both Riemann surfaces $\Sigma_{1,2}$ are non-zero.
In this case the manifold $M$ has $H^{2,0} (M) \neq 0.$
Let $E_{mn}$ be a non-trivial holomorphic $(2,0)$ form.
For simplicity we shall also assume that $\Sigma_1 =T^2 .$

In this case it is possible to introduce a superpotential into our model.
Let us consider the model with a quasihomogeneous superpotential
$W(x)= \lambda x^{N+1}/(N+1) ,$ where $\lambda$ is a coupling
constant (from now on we shall suppress $\lambda$ for simplicity).
In this case the $W_{1+\infty}$ algebra turns out to be
broken to a Virasoro one.

Using the equations of motion for the chiral supermultiplet (in the
presence of the superpotential)
it is easy to check that for any $N$ there is a unique bilinear \op \
which is $Q$-closed and holomorphic in $Q$-cohomology.
It has the following form
\beq
T_N=W_2 - \frac{1}{N+1} \partial_1 W_1 ,
\eeq
so that
\beq
\{ Q,T_N \} =0, \;\;\; \partial_{\bar{1}} T_N = \{ Q,...\} .
\eeq
Notice that the 1-component of the $U(1)$ current does not belong
to the $Q$-cohomology since the phase symmetry is broken by the
superpotential.

The \op \ $T_N$ has a spin 2 and generates the holomorphic Virasoro
algebra on $\Sigma_1 .$
In order to check it we calculate the OPE for $T_N (z,\bar{z})
T_N (w,\bar{w}) .$
It can be done in the weak coupling limit similarly with the calculation
in Ref.\cite{WiLG,SiWi}.
The point is that the superpotential terms in the Lagrangian are
dimensionful and make less singular contributions to the OPE as compared
to the free ones.
In particular it is easy to calculate the central charge which is given by
\beq
c_N= \chi \left( 1-\frac{6}{N+1} +\frac{6}{(N+1)^2} \right) .
\eeq
The model is renormalizable only if $N=1,2.$
For the case $N=1$ which corresponds to the free massive model
we have $c_2=-\chi/2,$ while for $N=2$ $c_3 =-\chi/3.$
In both cases the central charge is non-negative since we assumed
that $g>0 .$

Let us now discuss the behaviour of this algebra with respect to the
renormalization group.
The superpotential is not $Q$-exact.
Hence the physical correlators can depend on the coupling constant in
the superpotential.
This is a difference of the present case from the twisted $N=1$
Yang-Mills theory \cite{ajoh} where the whole action is $Q$-exact.
However the superpotential is not renormalizable
\cite{book} while the $D$-terms which are $Q$-exact are the
only renormalizable ones.
Thus we conclude that our construction is invariant under the
renormalization group.
This fact can also be understood as follows.
The quantum effects result in only a renormalization of the wave
functions by a factor $Z.$
In turn the same factor $Z$ appears in the \op \ $T_N$ due to quantum
effects.
These effects are non-trivial for the \op \ $T_N$ because it is
$Q$-closed only on mass-shell.
Notice that since the \op \ $T_N$ is an integral of a component of the
conserved energy-momentum tensor it does not acquire any its own
renormalization factor.
After a redefinition of the quantum fields the factor $Z$ disappears
in the OPE for $T_N$ which is therefore invariant under the
renormalization group.

Let us now consider the heterotic topological gauge theory.
Let us introduce the following gauge invariant operator
\beq
T=2 \pi \sg2 d^2 u \left(
\frac{1}{e^2}[2g^{2\bar{2}} F_{12} F_{1\bar{2}} - i (D_1
\bar{\lambda}) \chi_1 ] +
[ -(D_1\bar{\phi})(D_1 \phi) +
g^{2\bar{2}} \bar{\psi}_{12} (D_1 \psi_{\bar{2}}) ] \right) .
\eeq
This \op \ $T (z,\bar{z})$ obeys
\beq
\{ Q,T\} =0 ,\;\;\; \partial_{\bar{1}} T = \{ Q, ...\}.
\eeq
Thus the operator algebra generated by $T$ is holomorphic in the
$Q$-cohomology \footnote{One could suspect that there exists an anomaly in
eq.(4.13) similar to that of ref.\cite{SiWi}.
It is easy however to check that this is not the case provided that
the mixed anomaly in the effective action gets cancelled \cite{joh}.}.
Fixing the gauge in the Lagrangian (2.1)
one can calculate the \op \ product expansion in the weak coupling
constant limit because the action of the theory is $Q$-exact.
It is easy to check that the \op \ $T$ indeed generates the Virasoro
algebra with the central charge
\beq
c=({\rm dim} R - {\rm dim} G) \chi .
\eeq
This value of central charge agrees with the expression for the
gravitational anomaly (eq.(2.17)).

\section{Conclusions}
\setcounter{equation}{0}

We have shown that in a twisted $N=1$ SUSY model with a
single free chiral supermultiplet on the four manifold $M=\Sigma_1
\times \Sigma_2$ there exist two
infinite dimensional symmetries $W_{1+\infty}$ in the cohomology of the BRST
\op .
The generators of these algebras are integrals over $\Sigma_2$
($\Sigma_1$) of the bilinear
composite \op s.
The central charge of the $W_{1+\infty}$
algebra is the Euler characteristic $\chi_2$ ($\chi_1$) of
$\Sigma_2$ ($\Sigma_1$).
If we switch on a non-trivial quasihomogeneous
superpotential these algebras are reduced to
the chiral Virasoro algebras with central charges
proportional to $\chi_{1,2} .$
In the heterotic gauge theory there exist two chiral Virasoro algebras
with the central charges $\chi_{1,2} .$

Notice that one can interpret the chiral Virasoro algebra in $Q$-cohomology
as an algebra in a 2D conformal theory
on a surface $\Sigma_1$ with a classical metric.
It is tempting to try to extend such an interpretation allowing the
metric on $\Sigma_1$ to be a quantum one.
In this case we expect to get a 2D quantum gravity extracted from
the 4D theory.

It is worth noticing that the representation of $W_{1+\infty}$ given
here is very close to a representation of this algebra in terms of
2D free (fermionic or bosonic) fields \cite{bakas1,depireux,bakas}.
It is amusing however that in our representation the central charge
has a purely geometric origin.

We also point out that one can try to extend our construction to a model with
a multiple number of chiral supermultiplets.
In such a model it will be interesting to see if
a realization of $W_N$ and $W^p_{\infty}$ \cite{bakas1}
algebras can similarly be obtained
directly from a four dimensional field theory.

\section{Acknowledgments}

I am grateful to M.Axenides for interesting discussions of the results of
this work and a careful reading of the manuscript.
I also thank the high energy group at NBI where
this work was finished for its hospitality.
The present work was supported in part by a NATO grant GRG 930395.

\end{document}